\documentclass{jps-cp}
\usepackage{txfonts} 
\usepackage{times, color}

\title{Hadron and Nuclear Physics on the Light Front}

\author{Stanley J. \textsc{Brodsky}$^{1}$ and Alexandre \textsc{Deur}$^{2}$}

\inst{$^{1}$SLAC National Accelerator Laboratory, Stanford University,
Stanford, California 94309, USA\\
$^{2}$Thomas Jefferson National Accelerator Facility, Newport News, VA 23606, USA}

\email{sjbth@slac.stanford.edu, deurpam@jlab.org}

\recdate{July 31, 2019}

\abst{
The QCD light-front (LF) Hamiltonian equation $H_{LF}|\Psi\rangle = M^2 |\Psi\rangle$ derived from quantization 
at fixed LF time $\tau = t+z/c$ provides a causal, frame-independent, method for computing hadron 
spectroscopy as well as dynamical observables such as structure functions, transverse momentum 
distributions, and distribution amplitudes. The LF formalism also leads to  novel nuclear 
phenomena, such as  ``hidden color'', ``color transparency'', ``nuclear-bound quarkonium" and 
the  shadowing  and antishadowing of nuclear structure functions.  
For example, there are five distinct color-singlet Fock state representations of the six color-triplet 
quarks of the deuteron.  The hidden color Fock states become manifest  when the deuteron is 
probed when it has small transverse size, as in measurements of the deuteron form factor at large momentum transfer.     
The QCD Lagrangian with zero quark mass has no explicit mass scale. However, as shown by
de Alfaro, Fubini, and Furlan (dAFF), a mass scale can appear in the equations of motion without 
affecting the conformal invariance of the Action. When one applies the dAFF procedure to the QCD 
LF Hamiltonian, it leads  to a color confining potential $\kappa^4 \zeta^2$ for mesons, where 
$\zeta^2$ is the LF radial variable conjugate to  the $q \bar q$ invariant mass squared.  The same 
result, including spin terms, is obtained using LF holography (AdS/QCD)  -- the duality 
between LF dynamics and AdS$_5$ --  if one  
modifies the AdS$_5$ Action by the dilaton $\exp{+\kappa^2 z^2}$ in the fifth dimension $z$.   
If  this procedure is generalized using superconformal algebra, the resulting LF eigensolutions 
provide unified Regge spectroscopy of mesons, baryons, and tetraquarks, including remarkable 
supersymmetric relations between the masses of mesons, baryons and tetraquarks with  a universal Regge slope. 
One also obtains empirically viable predictions for spacelike and timelike hadronic form factors, 
structure functions, distribution amplitudes, and transverse momentum distributions. AdS/QCD 
also predicts the analytic  form of the nonperturbative running coupling  
$\alpha_s(Q^2) \propto \exp{(-{Q^2/4 \kappa^2}})$,  in agreement with the effective charge  
measured from measurements of the Bjorken sum rule.  The mass scale $\kappa$ underlying 
hadron masses  can be connected to the parameter   $\Lambda_{\overline {MS}}$ in the QCD 
running coupling by matching the nonperturbative dynamics to the perturbative QCD  regime. 
The result is an effective coupling $\alpha_s(Q^2)$  defined at all momenta.}

\kword{Light Front, Holography, Superconformal Algebra,Hadronic Spectroscopy, Nuclear Physics, 
running coupling, structure functions, form factors\ldots}

\begin{document}
\maketitle

\section{Introduction}
One of the most powerful theoretical tools for the application of quantum chromodynamics (QCD) to hadron 
physics is ``light-front'' (LF) quantization, based on the use of Dirac's ``front-form'' time 
$\tau = t+z/c$~\cite{Dirac:1949cp}, where $(t,\vec x)$ are the usual space-time coordinates. 
The LF wavefunctions (LFWFs) of hadrons and nuclei are the eigensolutions of the QCD 
Hamiltonian $H_{LF}|\Psi_H \rangle = M^2 |\Psi_H\rangle$, where $M$ is the mass of the state $\Psi_H$. 
Unlike ordinary Hamiltonian field theory quantized at fixed ``instant time'' $t$, the LFWFs of the eigenstates 
$\Psi_{n/H}(x_i, \vec k_\perp i, \lambda_i) = \langle\Psi_H|n\rangle$   are causal and Poincar\'e invariant; i.e., they are 
independent of the hadron's relative motion. Here $x_i = {k^+\over P^+ }=  {k^0 + k^3\over P^0+P^3}$ is 
the boost-invariant momentum fraction of a constituent in an $n$-particle Fock state $|n\rangle$. 
There is no physical Lorentz contraction of a bound-state in motion since the LFWFs
do not depend on the choice of Lorentz frame. 

As first noted by Terrell~\cite{Terrell:1959zz} and Penrose~\cite {Penrose:1959vz} (see also Weisskopf~\cite{Weisskopf}), 
the Lorentz contraction of a moving object is unobservable. 
For instance, the structure of a proton measured in deep inelastic lepton-proton scattering $e p \to e^\prime X$,  does 
not depend on whether it is measured with the proton frame a rest as in fixed-target experiments, or in motion as
in an electron-proton collider.  Factorization theorems such as the Drell-Yan  
cross-section for lepton pair production are independent of the observer's Lorentz frame. 
Lorentz contraction could only apply if one could observe an object  of a finite size at a single fixed time $t$, 
but this  boundary condition violates causality.

LF quantization is the natural formalism for relativistic quantum field theory since one obtains
a fundamental, boost-invariant dynamical description of bound-states.  
Measurements of hadron structure, such as deep inelastic lepton-proton scattering, are made at 
fixed LF time $\tau$, analogous to a flash photograph, not at a single ``instant time'' $t$.
Virtually every observable used in particle and nuclear physics is based on LF quantization. 
Structure functions are the squares of LFWFs, and form factors are computed from the overlaps of LFWFs. 
They also underly ``hadronization'', the conversion of quark and gluon quanta to a hadron at the amplitude level.
The boost invariance of the LFWFs ensures that measurements in the center of mass (CM) frame, 
e.g. for an electron-ion collider, give the same results as measurements in the target rest frame. 

As Dirac emphasized~\cite{Dirac:1949cp}, a Lorentz boost in the ``instant-form'' formalism 
necessarily introduces a boost operator with mixed kinematical and dynamical dependence.
In fact,  in the ordinary instant-form formalism, dynamical corrections to the boosted wavefunction (WF) of a moving bound-state are 
necessary to compute form factors, Compton amplitudes, and other observables.  
The boost of the instant-form WF cannot be resolved  within the framework of the instant-form alone.
For example, the boost of the deuteron instant-form WF is not 
the product of separate Wigner (or Melosh) boosts~\cite{Wigner:1939cj} of the proton and neutron.
Consequently, an apparently {\it ad-hoc}  spin-orbit correction is needed in order to restore the low 
energy theorem (LET)~\cite{Low:1954kd} since even an infinitesimal boost in the instant-form involves 
contributions which depend on the internal dynamics. Similarly,  the derivation of the 
Gerasimov-Drell-Hearn (GDH) sum rule~\cite{Gerasimov:1965et,Drell:1966jv,Hosoda:1966} fails 
unless one computes the matrix elements of the electromagnetic current  using the correct, dynamically 
boosted instant-form WF of the target~\cite{Brodsky:1968xc}.  
The required boost for atomic systems in quantum electrodynamics (QED) can be derived 
by adopting a  covariant approach such as the 
Bethe-Salpeter bound-state formalism, but this approach is outside the instant-form framework;  thus the effects of 
the dynamical contributions to the boost in the instant-form have to be parametrized as a ``fictitious force''  -- in the sense that it is 
an inertial force rather than of fundamental origin.
The problem of constructing the boosted instant-form WF is more complex in QCD, since
the required boost of a hadronic or nuclear bound-state combines
non-perturbative dynamics with kinematics.  The boost of a bound-state WF in QCD is generally
not calculable in the instant-form, which in turn makes an analytic analysis intractable.

There is a further complication  when one computes the form factors and other observables for bound-states using the instant-form.  
The evaluation of form factors and other matrix elements of currents
cannot be computed just from 
the integrated overlap
of the instant-form WFs since there are additional  contributions from the current of pairs of charged constituents 
which arise from the vacuum.  
For example, in QED one must couple the external current 
$J^\mu$ to the leptons which are created or annihilated from the vacuum 
by the $e^+ e^- \gamma $ interaction. These ``vacuum-induced'' contributions to the current 
appear in the instant-form 
when one relates Feynman diagrams to the sum of ``old-fashioned'' 
time-ordered perturbation theory diagrams, as in Wick's theorem.  The correct Poincar\'e invariant 
results for form factors cannot be obtained in the instant-form without these contributions. 
In contrast, they do not appear in the front-form since all particles have positive $k^+$. 

\section{Novel QCD Nuclear Phenomena}

The LF formalism leads to many novel nuclear phenomena, such as  ``hidden color''~\cite{Brodsky:1983vf}
``color transparency"~\cite{Brodsky:1988xz},  ``nuclear-bound quarkonium"~\cite{Brodsky:1989jd}, ``nuclear shadowing and antishadowing'' of nuclear structure functions, etc.  For example, there are five distinct color-singlet Fock state representations of the 
six color-triplet quarks of the deuteron.  These hidden-color Fock states become manifest  when 
the deuteron fluctuates to a small transverse size,
as in measurements of the deuteron form factor at 
large momentum transfer. One can also probe the hidden-color Fock states of the deuteron by 
studying the final state of the diffractive dissociation of the deuteron on a nucleus  $D A \to X A$, 
where $X$ can be $\Delta^{++} +\Delta^-$,  six quark jets, or other novel final states.  

More generally, one can measure the LFWFs of a hadron by using the ``Ashery method''~\cite{Aitala:2000hb}:
in the diffractive dissociation of a high energy hadron into quark and gluon jets by two-gluon exchange,  
the cross-section measures the  square of the second transverse derivative of the projectile LFWF. 
Similarly, the dissociation of a high energy atom such as positronium or ``true muonium'' ($[\mu^+\mu^-]$) 
can be used to measure the transverse derivative of its LFWFs.

LFWFs provide the input for scattering experiments at the amplitude level, 
encoding the structure of a projectile at a single LF time $\tau$.  For example, consider photon-ion 
collisions. The incoming photon probes the finite size structure of  the incoming nucleus at fixed LF time, like 
a photograph -- not at a fixed instant time, which is acausal. 
Since the nuclear state is an eigenstate of the LF Hamiltonian, its structure is independent of its momentum, 
as required by Poincar\'e invariance.  One gets the same answer in the ion rest frame, the CM frame, or even if 
the incident particles move in the same direction, but collide transversely.  There are no colliding ``pancakes'' using the LF formalism. 

The resulting photon-ion cross-section is not point-like; it is shadowed: 
$\sigma(\gamma A \to X) = A^\alpha \sigma(\gamma N \to X)$, 
where  $A$ is the mass number of the ion, $N$ stands for a nucleon, and the power  $ \alpha \approx 0.8$  
reflects Glauber shadowing.  The shadowing stems from the destructive 
interference of two-step and one-step amplitudes, where the two-step processes involve diffractive reactions  
on a front-surface nucleon which shadows the interior nucleons. Thus the photon interacts primarily on the 
front surface. Similarly  a high energy ion-ion collision  $A_1 + A_2 \to X$ involves the overlap of the incident
frame-independent LFWFs. The initial interaction on the front surface of the colliding ions can resemble a shock wave.

In the case of a deep inelastic lepton-nucleus collision $\gamma^* A \to X$, the two-step amplitude 
involves a leading-twist diffractive deep inelastic scattering (DDIS) 
$\gamma^* N_1\to V^* N_1$  on a front 
surface nucleon $N_1$ and then the on-shell propagation of the vector system $ V^*$ to a downstream 
nucleon $ N_2$ where it interacts inelastically: $V^* N_2 \to X$.  If the DDIS involves Pomeron exchange, 
the two-step amplitude  interferers destructively  with the one-step 
amplitude  $\gamma^* N_1 \to X$ thus producing shadowing of the nuclear parton distribution function at low $x_{bj} < 0.1$
where $x_{bj}$ is the Bjorken scaling variable.  
On the other hand, if the DDIS process involves $I=1$ Reggeon exchange, the interference is constructive, 
producing {\it flavor-dependent}  leading-twist antishadowing ~\cite{Brodsky:1989qz}  in the domain $0.1 < x_{bj} < 0.2$.

\section{Light-front Holography} 
 
Five-dimensional AdS$_5$ space provides a geometrical representation of the conformal group.  
Remarkably,  AdS$_5$  is holographically dual to $3+1$  spacetime at fixed LF time $\tau$~\cite{Brodsky:2014yha}.  
A color-confining LF  equation for mesons of arbitrary spin $J$ can be derived~\cite{deTeramond:2013it}
from the holographic mapping of  the ``soft-wall model'' modification of AdS$_5$ space for the specific dilaton 
profile $e^{+\kappa^2 z^2}$, where $z$ is the  fifth dimension variable of  the five-dimensional AdS$_5$ space. 
A holographic dictionary  maps the fifth dimension $z$  to the LF radial variable $\zeta$,  with $\zeta^2  =  b^2_\perp(1-x)$.   
The same physics transformation maps the AdS$_5$  and $(3+1)$ LF expressions for electromagnetic and 
gravitational form factors to each other~\cite{deTeramond:2013it}.

\begin{figure}
 \begin{center}
\includegraphics[height=12cm,width=15.2cm]{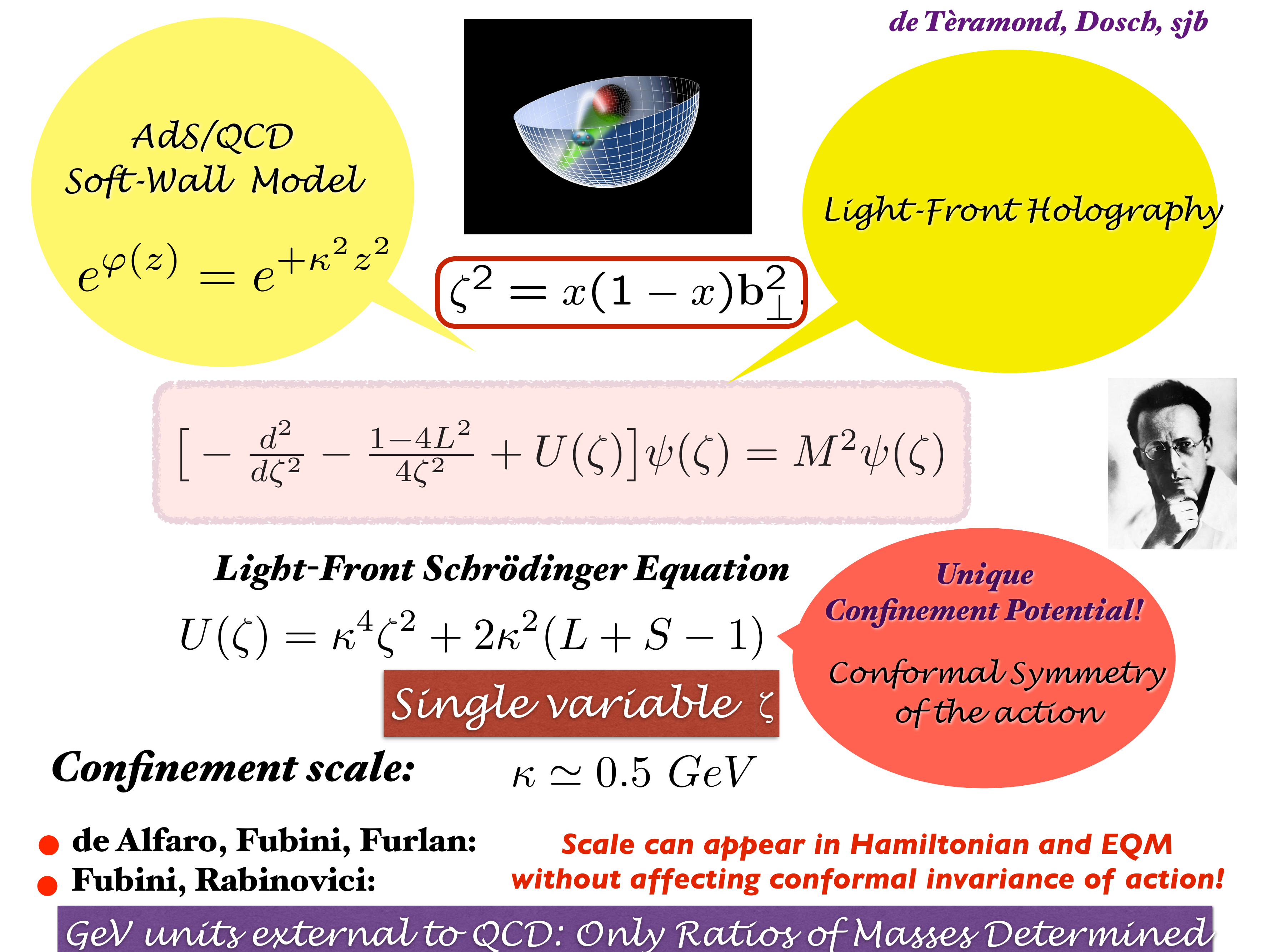}
\end{center}
\caption{The LF Schr\"odinger Equation derived from the LF Holography and the AdS/QCD correspondence. When one applies the dAFF procedure to the QCD 
LF Hamiltonian, it leads  to a color confining potential $\kappa^4 \zeta^2$ for mesons, where 
$\zeta^2$ is the LF radial variable conjugate to  the $q \bar q$ invariant mass squared.  The same 
result, including spin terms, is obtained using LF holography  -- the duality 
between LF dynamics and AdS$_5$ --  if one  
modifies the AdS$_5$ action by the dilaton $\exp{+\kappa^2 z^2}$ in the fifth dimension $z$.   
\label{NN2018Summary}}
\end{figure}

A key tool is the remarkable de Alfaro, Fubini, and Furlan (dAFF)~\cite{deAlfaro:1976je} principle which 
shows how a mass scale can appear in a Hamiltonian and its equations of motion while retaining the 
conformal symmetry of the action.  When one applies the dAFF procedure to LF holography, a mass 
scale $\kappa$ appears which determines universal Regge slopes, and the hadron masses.
The  resulting ``LF Schr\"odinger Equation''   incorporates color confinement and other 
essential spectroscopic and dynamical features of hadron physics, including Regge theory~\cite{the:Regge}, 
the Veneziano formula~\cite{Veneziano:1968yb}, a massless pion for zero quark mass and linear Regge trajectories with the 
universal slope  in the radial quantum number $n$   and the internal  orbital angular momentum $L$.  
The combination of LF dynamics, its holographic mapping to AdS$_5$ space, and the 
dAFF procedure provides new insight into the physics underlying color confinement, the 
nonperturbative QCD coupling, and the QCD mass scale.  The $q \bar q$ mesons and their 
valence LFWFs are the eigensolutions of the frame-independent a relativistic bound-state LF 
Schr\"odinger equation. The formalism is summarized in Fig.~\ref{NN2018Summary}.
The mesonic $q\bar  q$ bound-state eigenvalues for massless quarks are $M^2(n, L, S) = 4\kappa^2(n+L +S/2)$.
This equation predicts that the pion eigenstate  $n=L=S=0$ is massless for zero quark mass. 
When quark masses are included in the LF kinetic energy $\sum_i  {k^2_{\perp i} + m^2\over x_i}$,  the  spectroscopy of  mesons  are 
predicted correctly, with equal slope in the principal quantum number $n$ and the internal orbital angular momentum $L$.   %
 A comprehensive review is given in  Ref.~\cite{Brodsky:2014yha}. 
 
\section{Superconformal Algebra and Hadron Physics} 
If one generalizes LF holography using {\it superconformal algebra}
the resulting LF eigensolutions yield 
a unified Regge spectroscopy of mesons, baryons and tetraquarks, including remarkable supersymmetric 
relations between the masses of mesons and baryons of the same parity~\cite{Brodsky:2013ar,Brodsky:2016nsn}.   
LF Holography, together with superconformal algebra, not only predicts meson and baryon  spectroscopy 
consistent with measurements, but also hadron dynamics;  this includes  vector meson electroproduction,  
hadronic LFWFs, distribution amplitudes, form factors, and valence structure functions.  Applications to the 
deuteron elastic form factors and structure functions is given in Refs.~\cite{Gutsche:2015xva,Gutsche:2016lrz}

\begin{figure}
 \begin{center}
\includegraphics[height=12cm,width=15.2cm]{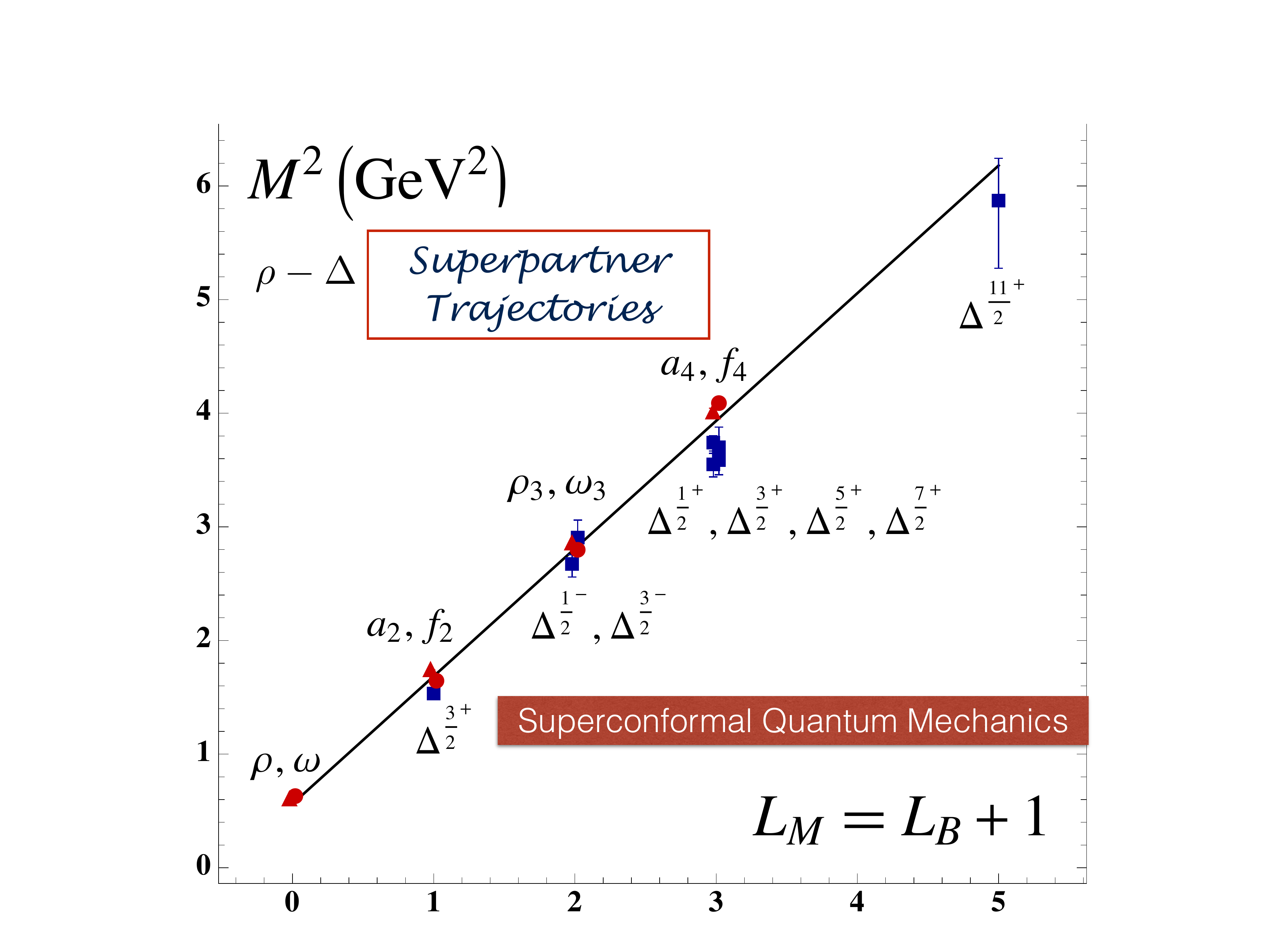}
\end{center}
\caption{Comparison of the $\rho/\omega$ meson Regge trajectory with the $J=3/2$ $\Delta$  baryon trajectory. 
Superconformal algebra  predicts the mass  degeneracy of the meson and baryon trajectories if one identifies a 
meson with internal orbital angular momentum $L_M$ with its superpartner baryon with $L_M = L_B+1.$
See Refs.~\cite{deTeramond:2014asa,Dosch:2015nwa}. 
\label{NSTARFigB}}
\end{figure}

QCD is not supersymmetrical in the usual sense --the QCD Lagrangian is based on quark and gluonic 
fields-- not squarks or gluinos. However, its hadronic eigensolutions conform to a representation of superconformal 
algebra, reflecting the underlying conformal symmetry of chiral QCD and its Pauli matrix representation.   
A comparison with experiment is shown in Fig.~\ref{NSTARFigB}.

The eigensolutions of superconformal algebra predict the Regge spectroscopy of mesons, baryons, and 
tetraquarks of the same parity and twist as equal-mass members of the same 4-plet representation with 
a universal Regge slope.~\cite{Dosch:2015nwa, Brodsky:2016rvj, Nielsen:2018ytt},   The $q \bar q$ mesons with orbital angular momentum $L_M=L_B+1$ have 
the same mass as their baryonic partners with orbital angular momentum $L_B$~\cite{deTeramond:2014asa,Dosch:2015nwa}.

The predictions from LF holography and superconformal algebra can also be extended to mesons, baryons, 
and tetraquarks with strange, charm and bottom quarks.  
Although conformal symmetry is strongly broken by the heavy quark masses, the basic
underlying supersymmetric mechanism, which transforms mesons to baryons (and
baryons to tetraquarks), still holds and gives remarkable mass degeneracy across the entire spectrum of light, 
heavy-light and double-heavy hadrons.
One also obtains viable predictions for spacelike and timelike hadronic form factors, structure functions, 
distribution amplitudes, and transverse momentum distributions~\cite{Sufian:2016hwn,deTeramond:2018ecg}.

\section {The QCD Running Coupling at all Scales} 

The QCD running coupling $\alpha_s(Q^2)$
sets the strength of  the interactions of quarks and gluons as a function of the momentum transfer $Q$.
The dependence of the coupling $Q^2$ is needed to describe hadronic interactions at 
both long and short distances~\cite{Deur:2016tte}. 
The QCD running coupling can be defined~\cite{Grunberg:1980ja} at all momentum scales from a 
perturbatively calculable observable, such as the coupling $\alpha_s^{g_1}(Q^2)$, which is defined 
using the Bjorken sum rule~\cite{Bjorken:1966jh}, and determined from the sum rule prediction at high $Q^2$ and, below, from its 
measurements~\cite{Deur:2004ti,Deur:2014vea,Deur:2008ej}.  At high $Q^2$, such ``effective charges'' 
satisfy asymptotic freedom, obey the usual pQCD renormalization group equations, and can be related 
to each other without scale ambiguity by commensurate scale relations~\cite{Brodsky:1994eh}.  

\begin{figure}
\begin{center}
\includegraphics[height=10cm,width=14cm]{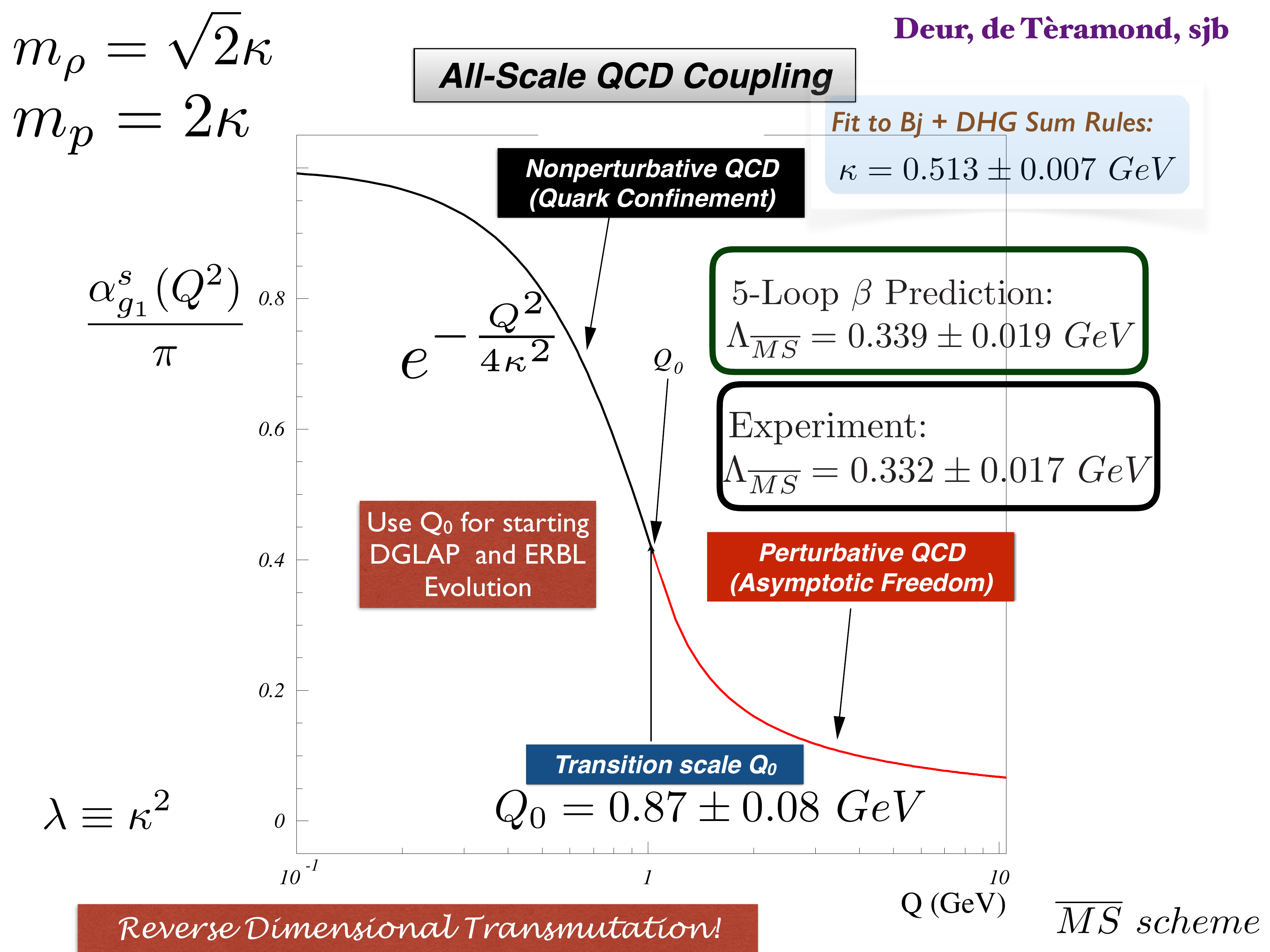}
\end{center}
\caption{
Prediction from LF Holography and pQCD for the running coupling $\alpha_s^{g_1}(Q^2)$ at all scales.   
The magnitude and derivative of the perturbative and nonperturbative coupling are matched at the scale $Q_0$.  
This matching connects the perturbative scale 
$\Lambda_{\overline{MS}}$ to the non-perturbative scale $\kappa$ which underlies the hadron mass scale. 
\label{DeurCoupling}}
\end{figure}

The dilaton  $e^{+\kappa^2 z^2}$ soft-wall modification of the AdS$_5$ metric, 
together with LF holography, predicts the functional behavior of the running coupling
in the small $Q^2$ domain~\cite{Brodsky:2010ur}: ${\alpha_s^{g_1}(Q^2) = \pi   e^{- Q^2 /4 \kappa^2 }}$. 
Measurements of  $\alpha_s^{g_1}(Q^2)$~\cite{Deur:2005cf,Deur:2008rf} are remarkably consistent  
with this predicted Gaussian form; the best fit gives $\kappa= 0.513 \pm 0.007$~GeV, see Fig.~\ref{DeurCoupling}.
We have also shown~\cite{Brodsky:2010ur,Deur:2014qfa,Brodsky:2014jia} with de T\'eramond how the parameter $\kappa$,  
which  determines the mass scale of  hadrons and Regge slopes  in the zero quark mass limit, can be 
connected to the  mass scale $\Lambda_s$  controlling the evolution of the perturbative QCD coupling.  
The high $Q^2$ dependence  of $\alpha_s^{g_1}(Q^2)$ is  predicted  by  pQCD.  
The matching of the high and low $Q^2$ regimes of $\alpha_s^{g_1}(Q^2)$ -- both its value and 
its slope -- then determines a scale $Q_0$ which sets the interface between perturbative 
and nonperturbative hadron dynamics. In the $\overline{MS}$ scheme, $Q_0 =0.87 \pm 0.08$ GeV. 
This connection can be done for any choice of renormalization scheme. 
The result of this perturbative/nonperturbative matching is an effective QCD coupling  defined at all momenta.   
The predicted value of $\Lambda_{\overline{MS}} = \kappa e^{-a}\sqrt{2/a}= 0.339 \pm 0.019$~GeV 
(with $a=4\big[\sqrt{\ln(2)^2+1+\beta_0/4}-\ln(2)\big]/\beta_0 + \mathcal{O}(\beta_1)$) from this analysis agrees well 
the measured value~\cite{Agashe:2014kda}  $\Lambda_{\overline{MS}} = 0.332 \pm 0.017$~GeV.
These results, combined with the AdS/QCD superconformal predictions for hadron spectroscopy, 
allow one to compute hadron masses in terms of $\Lambda_{\overline{MS}}$:
$m_p =  \sqrt 2 \kappa = 3.21\Lambda_{\overline{MS}}$,  $m_\rho = \kappa = 2.2\Lambda_{\overline{ MS} }$ 
and $m_p = \sqrt 2 m_\rho$, meeting a challenge proposed by Zee~\cite{Zee:2003mt}. The value of $Q_0$ 
can be used to set the factorization scale for DGLAP evolution of hadronic structure 
functions~\cite{Gribov:1972ri,Altarelli:1977zs,Dokshitzer:1977sg} and the ERBL evolution of distribution 
amplitudes~\cite{Lepage:1979zb,Efremov:1979qk}.
We have also computed the dependence of $Q_0$ on the choice of the effective charge used to define 
the running coupling and on the renormalization scheme used to compute its behavior in the perturbative regime.   
The use of  the scale $Q_0$  to  resolve  the factorization scale uncertainty in structure functions and fragmentation 
functions,  in combination with the scheme-independent {\it principle of maximum conformality} 
(PMC )~\cite{Mojaza:2012mf} for  setting renormalization scales,  can 
greatly improve the precision of pQCD predictions for collider phenomenology.

\section{Summary}
The new approach to Poincar\'e invariant hadron dynamics based on light-front holography discussed here has many attractive features, 
including color confinement, the analytic form of the LF confinement potential,  a massless quark-antiquark pion bound state in the chiral limit, 
The incorporation of supersymmetric algebra  leads to  4-plets of mass degenerate $q \bar q$ mesons, quark-diquark baryons and diquark/antidiquark tetraquarks.
All of the Regge Trajectories have universal slopes in $n$ and $L$. The resulting causal boost-invariant LF wavefunctions predict hadronic form factors, structure functions, transverse momentum distributions, and other dynamical observables such as counting rules and hadronization at the amplitude level.  
We also obtain the form of the QCD coupling at all scales, consistent with experiment, thus 
connecting perturbative and nonperturbative hadron dynamics.

\section*{Acknowledgments}
The physics results reported here are based on work developed with our collaborators, including Guy de T\'eramond,  H. Guenter Dosch, Marina Nielson, Cedric Lorc\`e, Josh Erlich, and R.S. Sufian.
This work is based in part upon work supported by the U.S. Department of Energy, the Office of Science, 
and the Office of Nuclear Physics under contract DE-AC05-06OR23177.  This work is also supported by the 
Department of Energy contract DE--AC02--76SF00515.   SLAC-PUB 17464.

\end{document}